\documentstyle[12pt]{article}
\def\simlt{\stackrel{<}{{}_\sim}}
\def\simgt{\stackrel{>}{{}_\sim}}
\def\be{\begin{equation}}
\def\ee{\end{equation}}
\def\bear{\be\begin{array}}
\def\eear{\end{array}\ee}
\def\bea{\begin{eqnarray}}
\def\eea{\end{eqnarray}}

\def\baselinestretch{1}
\begin{document}
\catcode`@=11
\newtoks\@stequation
\def\subequations{\refstepcounter{equation}%
\edef\@savedequation{\the\c@equation}%
  \@stequation=\expandafter{\theequation}
  \edef\@savedtheequation{\the\@stequation}
  \edef\theequation{\theequation}%
  \setcounter{equation}{0}%
  \def\theequation{\theequation\alph{equation}}}
\def\endsubequations{\setcounter{equation}{\@savedequation}%
  \@stequation=\expandafter{\@savedtheequation}%
  \edef\theequation{\the\@stequation}\global\@ignoretrue

\noindent}
\catcode`@=12
\begin{titlepage}

\title{{\bf Conditions for Viable Affleck--Dine
Baryogenesis--} \break {\bf Implications for String Theories}
\thanks{Research supported in part by: the CICYT
(contract AEN95-0195) and the European Union
(contract CHRX-CT92-0004) (JAC); 
and the US Department of Energy
under grant DE-FG03-91ER40662 Task C (GG).}
}
\author{ {\bf J. A. Casas
${}^{ {\footnotesize\P}}$},
and {\bf G. B. Gelmini${}^{ {\footnotesize\S}}$}\\
\hspace{3cm}\\
${}^{\footnotesize\P}$ {\small Instituto de
Estructura de la Materia, CSIC}\\
{\small Serrano 123, 28006 Madrid, Spain}\\
{\small casas@cc.csic.es}\\
${}^{\footnotesize\S}$ {\small UCLA
 (University of California, Los Angeles)}\\
 {\small 405 Hilgard Ave. CA 90095, USA}\\
{\small gelmini@physics.ucla.edu}}
\date{}
\maketitle

\def\baselinestretch{1.15}
\begin{abstract}
\noindent
We examine the conditions for  a viable Affleck--Dine baryogenesis 
in supergravity (SUGRA) scenarios, finding surprisingly strong
constraints on the type of SUGRA theory. These constraints are 
beautifully fulfilled by  string-based SUGRA models provided that 
inflation is driven by a modulus ($T$) field.

\end{abstract}

\thispagestyle{empty}

\vspace{3cm}
\vskip-20cm
\rightline{}
\rightline{ IEM-FT-157/97}
\rightline{ UCLA/97/TEP/15}
\rightline{ June 1997}
\rightline{hep-ph/9706439}

\vskip3in

\end{titlepage}
\newpage
\setcounter{page}{1}

\section{Introduction}

 The fundamental problem of how a net baryon number has been generated
in the universe is still far from being solved. The hopes that a
sufficient baryogenesis could be produced at the electroweak transition
of the Standard Model have been  proven not to be viable
 \cite{PT,CP}, which clearly 
points out to the need of new physics. Weak scale baryogenesis has also been 
studied in the context of supersymmetric models \cite{MSSM}. The results are
somewhat better than in the Standard Model, but a working scenario would
require a substantial amount of artificial arrangement of the 
supersymmetric parameters \cite{MSSM2}.

Given this situation it is important to study efficient alternative
mechanisms for baryogenesis, the most promising of which is
probably the one of Affleck and Dine (AD)  \cite{AD}. This is based on the
possible production in the early universe of large vacuum expectation 
values (VEV's) of fields (or combinations of fields) carrying baryon or 
lepton number ($B$ or $L$). If, in addition, 
$B$ or $L$ is explicitly broken
by some of these condensates, it is possible to excite a net baryon
number in the universe. This mechanism is remarkably efficient; as
a matter of fact, it tends to generate too much $B$, which needs to
be subsequently diluted. The AD mechanism
requires scalars carrying $B$ or $L$, which is happily the case of
supersymmetric (SUSY) theories. The aim of this paper is to establish
which conditions are to be fulfilled by a SUSY theory (more precisely,
by a supergravity (SUGRA) theory) to be consistent with AD baryogenesis,
and apply the results to relevant models (particularly string models).

Let us briefly review the basics of the AD mechanism. As mentioned
above, the AD mechanism requires large VEV's of scalar fields (say AD
fields) in the
early universe. These have been commonly associated in the
literature with the existence of (approximately) flat directions
involving the AD fields. Then quantum fluctuations during inflation 
may yield initial large VEV's, which are eventually driven to zero 
by e.g. low-energy SUSY breaking mass terms. This was the implementation
of the AD mechanism originally considered by Affleck and Dine. As has been
stressed  in Ref. \cite{DRT}, this picture ignores the following
basic fact.  During inflation
SUSY is necessarily spontaneously broken since the scalar potential
$V$ gets a VEV, $\langle V \rangle= 3 H^2 M_{P}^2$, and effective 
SUSY soft breaking terms, in particular effective soft 
masses of the order of the Hubble constant $H$
are generated ($M_P=M_{Planck}/\sqrt{8\pi}$ is the usual SUGRA Planck
scale).  
These spoil
the flat directions (although we will maintain this denomination), 
since $H$ is expected to be
 $O(10^{13-14}$ GeV) in order to obtain  the observed magnitude of 
 the microwave background radiation anisotropy in most  models.
However, large VEV's are still possible if these
effective masses squared are negative. 
In Ref. \cite{DRT} it was pointed out that this is
perfectly possible in a generic SUGRA theory. Then, the sequence of
events is schematically as follows.

\begin{itemize}

\item During inflation an effective potential for the AD field(s), say
$\phi$, is produced by the $O(H)$ soft terms. These must include negative
soft mass and, possibly, generalized A-type terms.
 For large values of $\phi$ the
flat direction of the potential must be lifted, which is likely done by 
F--terms coming from $\sim \phi^n$ terms in the
superpotential $W$. The $\phi$ field evolves rapidly towards its minimum
during this period.

\item After inflation the inflaton starts oscillating coherently
about its minimum. The energy density in these oscillations, which
 dominates the universe, red-shifts as matter so that  
$H= (2/3t)$ and the position of the minimum
of $V$ in the $\phi$  direction becomes time-dependent (through $H$). 
The $\phi$ field follows the
instantaneous minimum provided that the operators that lift the
flat direction of the 
potential are non-renormalizable (i.e. $n\ge 4$ in the previous
paragraph), which is perfectly possible.

\item At $H=O($TeV) the low-energy SUSY soft breaking  terms become
important. They produce a positive mass term for $\phi$ and 
$B$($L$)--violating A--type terms, which become comparable to the other
terms in the potential. For arbitrary phases of the $\phi$ fields 
the A--terms can also generate CP violation.
 In this period a large (maximal) $B$ or $L$ number is generated.
Subsequently the decay of the inflaton partially dilutes the asymmetry
yielding the final value of $B$. It turns out that for $n=4$ the latter
can be very naturally the observed one, though other values of $n$ 
(preferably $n\ge 4$) are also possible.

\end{itemize}

{}From the above picture it is clear that a successful AD baryogenesis
requires that the $\phi$ effective potential during inflation contains

\begin{enumerate}

\item negative effective mass terms, $m^2_\phi\le 0$,

\item non-renormalizable terms to lift flat directions of the potential.

\end{enumerate}

The purpose of the present paper is precisely to show what kinds of SUGRA
theories are consistent with these conditions. We will find surprisingly
strong constraints, which select a class of SUGRA theories. As an
additional
surprise, the possibility $m_\phi\simeq 0$, which had been considered 
implausible after the arguments of Ref.\cite{DRT}, 
turns out to be perfectly possible, allowing for  scenarios in
 which the AD mechanism can 
 be implemented in the original, ``old-fashioned", way. 

In Sect. 2 we examine the viability of AD baryogenesis in SUGRA
models when inflation (and the corresponding SUSY breaking) is
triggered by a non-vanishing D--term (D--inflation). Sect. 3
is devoted to the other possible case, i.e. F--inflation.
In Sect. 4 we apply our results to string scenarios,
finding very significant and encouraging results. Finally,
in Sect. 5 we summarize the conclusions.

\section{D--Inflation}

The possibility of D--inflation in SUGRA scenarios is very attractive
since, as has been often claimed in the literature, F--inflation seems
to lead naturally to too large inflaton mass terms that disable the
inflationary process (however, see footnote 4). 
Models of D--inflation were first proposed in 
Ref.\cite{CM} and have been recently revived in Ref.\cite{BD}.

In order to be concrete, it is convenient to
suppose
that inflation is mainly triggered by a single D--term (this does not
reduce the generality of the present analysis). Then, a suitable
choice is to suppose that the relevant D--term is associated to one
``anomalous'' $U(1)$, which takes the 
form\footnote{These ``anomalous'' $U(1)$
 symmetries appear  frequently in string theories\cite{DSW,CKM}. 
 The apparent anomaly is actually
cancelled by the transformation of the dilatonic axion. The form of
the corresponding D--term is that of a Fayet--Iliopoulos D--term.}
\be
V_D=\frac{1}{2}D^2=\frac{1}{2}g^2\left|
\xi+\sum_j q_j|z_j|^2 K_{j\bar j} \right|^2\;,
\label{FI}
\ee
where $g$ is the corresponding gauge coupling, $q_j$ are the charges of
all the chiral fields, $z_j$, under the anomalous $U(1)$ and 
$K_{j\bar j}$ is the K{\"a}hler metric, i.e. the second derivative
of the K{\"a}hler potential $K_{j\bar j}=\partial K/\partial z_j
\partial \bar z_j$ (we are assuming here a basis for the $z_j$ fields 
 where the K{\"a}hler metric is diagonal). Finally, the 
constant $\xi$
is related to the apparent anomaly, $\xi=g^2M_P^2(\sum_j
q_j/192\pi^2)$. This was precisely the scenario
considered in Refs.\cite{CM,BD}. At low energy the D--term is cancelled
by the VEV's of some of the scalars entering Eq.(\ref{FI}), but
initially $\langle D\rangle$ may be different from zero, thus triggering
inflation\footnote{E.g. in the model of the first article of 
Ref. \cite{BD} the combination of
the D--term (\ref{FI}) with appropriate F--terms provides nearly flat 
directions for  a slow rollover transition,
 in a sort of hybrid-like inflation. Then,
during the inflationary epoch, $(1/2)|\langle D\rangle|^2
= (1/2)g^2|\xi|^2$, i.e.  $H^2\sim (1/2)g^2|\xi|^2/M_P^2$.}.
Let us also notice that the scenario is quite insensitive to the
details of the K{\"a}hler potential $K$ (note in particular that
$(K_{j\bar j})^{1/2} z_j$ are simply the canonically normalized
chiral fields). In consequence we will ignore the K{\"a}hler factors,
taking $K_{j\bar j}=1$, in the rest of this section.

The AD mechanism requires the existence of a field $\phi$ 
(or several ones) carrying
$B$ or $L$ that develops a non-vanishing VEV during the inflationary
process, say $\langle\phi \rangle_{in}\neq 0$ (the subindex $in$ denotes
either inflation or initial). Experimental evidence requires the
final (low-energy) VEV to be vanishing, i.e.  $\langle\phi 
\rangle_f=0$. There are two main possibilities to consider, depending
on $q_\phi\neq 0$ or $q_\phi = 0$.

Let us start with the first case. Taking\footnote{If some fields
$\eta_j$ with $q_j\neq 0$ get  $\langle\eta_j \rangle_{in}\neq 0$,
we simply replace $\xi\rightarrow \xi+\sum_j q_j |\langle\eta_j
\rangle_{in}|^2$.} $(1/2)|\langle D\rangle|^2
\propto g^2|\xi|^2$, it is clear that the D--term induces an effective
mass term for $\phi$ whose sign depends on the relative sign of
$\xi$ and $q_\phi$. If 
\be
{\rm sign}(\xi)\  {\rm sign}(q_\phi) =-1, 
\label{signs}
\ee
then the
effective mass squared is {\em negative} and we expect $\langle\phi
\rangle_{in}\neq 0$.

However, this cannot be the whole story, since in the absence of
additional $\phi$--dependent terms in $V$, $\langle\phi
\rangle_{in}$ would adjust itself to cancel the D--term, thus disabling
the inflationary process and breaking $B$ or $L$ at low energy.
Thus, we need extra contributions yielding $\langle D
\rangle_{in}\neq 0$, $\langle\phi\rangle_f= 0$.  These may come from 

\begin{description}

\item[{\em a)}] low-energy soft breaking terms,

\item[{\em b)}] F--terms,

\item[{\em c)}] D--terms.

\end{description} 

\noindent

The first possibility, {\em a)}, cannot work in practice. Certainly, 
all the scalar fields  get soft-breaking masses, $m
=O($TeV), but they are too small to be useful here. Schematically,
the $\phi$--dependent potential reads
\be
V=\frac{1}{2}g^2\left|
\xi+ q_\phi|\phi|^2 \right|^2 + m^2|\phi|^2\;,
\label{VD}
\ee
which leads to  $|\langle\phi
\rangle_{in}|^2\simeq (-\xi/q_\phi)- (m^2 /g^2q_\phi^2)$ and
$\langle V \rangle_{in}\simeq (-m^2 /q_\phi)[\xi
+ (m^2 /2g^2q_\phi)]$, which is
too small to produce inflation. Even if, in order to preserve inflation,
$m$ were abnormally large, we notice from (\ref{VD}) that the potential
in this direction would be
 eventually lifted (for large $\phi$) by the quartic 
term $\sim g^2 q_\phi^2 |\phi|^4$. 
As mentioned in the previous section, such a term would imply that
the $\phi$--field does not follow the instantaneous minimum of $V$
after inflation. This may spoil the AD mechanism, that depends crucially
on the initial value of the field when it begins to oscillate
freely, at later times.

The  possibility {\em b)} is more plausible but still difficult to 
implement. In principle F--terms may contain masses much larger than 
$O($TeV), thus yielding a sizeable $\langle V \rangle_{in}$, suitable 
for inflation. However, this possibility is
still unattractive since the  problem of the field being driven away
from the instantaneous minimum remains. Nevertheless, it may well occur
that the relevant F--terms that lift the flat direction of the potential 
are not mass terms, but non-renormalizable terms, coming e.g. from a term
$\sim \lambda M^{-(n-3)}\phi^n$ ($n\ge 4$) 
in the superpotential
$W$, where $M$ is some large mass scale
and we take for convenience
$\lambda=O(1)$. Then, schematically, the potential reads
\be
V=\frac{1}{2}g^2\left|
\xi+ q_\phi|\phi|^2 \right|^2 + \frac{|\lambda|^2}{M^{2n-6}}
|\phi|^{2n-2}\;.
\label{VDF}
\ee
It is easy to check that in order to ensure that the non-renormalizable
term (and not the quartic one) be responsible for the lifting of the 
flat direction of the potential, it is necessary to have
$\xi/M^2 > 1$ (preferably much larger than 1). This is precisely the 
opposite to what one expects  in these models,
since non-renormalizable terms in SUGRA
are likely to be suppressed by inverse powers of $M_P$, and thus
$\xi/M^2 \ll 1$. Besides, it is straightforward to check that this would 
yield $|\langle\phi \rangle_{in}|^2\simeq (-\xi /q_\phi)$ and a
too small $\langle V(\phi)\rangle_{in}$ for inflation.

The  possibility {\em c)} arises if $\phi$ is charged under extra gauge
groups apart from the inflationary one, which is quite plausible.
The extra D--terms would induce in general quadratic and quartic terms
in $\phi$. For instance, a quartic term $\frac{1}{2}b|\phi|^4$, 
with $b=(g^2q_\phi^2)$, together
with the anomalous D--term (\ref{FI}), i.e.
\be
V=\frac{1}{2}g^2\left|
\xi+ q_\phi|\phi|^2 \right|^2 + \frac{1}{2}b|\phi|^4\;
\label{VDD}
\ee
would give $|\langle\phi \rangle_{in}|^2\simeq$
$-g^2 q_\phi\xi / ( g^2q_\phi^2 + b)< (-\xi / q_\phi)$,
$\langle V\rangle_{in} = O(g^2 \xi^2)$, 
which is perfectly consistent with inflation. 
The potential problems here are that the flat direction of the potential
is lifted by quartic terms, which is inescapable in this scenario, and
the absence of A--terms at high energy scales.
$B$ and $CP$ violating A--terms are required by the AD mechanism to drive
the generation of $B$.
These terms are absent in
the scenario  outlined above, but they can appear at $H \le O($TeV),
when the low-energy soft breaking terms become relevant.

\vspace{0.3cm}
\noindent

It is clear, therefore, that in D--inflationary scenarios,  
it is enough to have the AD field $\phi$  
 carry a charge of the appropriate sign under the $U(1)$ responsible 
for the inflation, in order to obtain
an effective negative mass squared at $\phi$=0 during inflation (and, 
consequently, 
$\langle\phi\rangle_{in}\neq 0$). However, the remaining details for
a successful AD mechanism (and D--inflation) are not so easy to arrange.

\vspace{0.3cm}
\noindent
Nevertheless, there is a beautiful alternative to get AD baryogenesis. 
Namely, if $q_\phi=0$ (i.e. the AD field(s) is (are) not charged under the 
inflationary $U(1)$), then $m_\phi^2=0$ during inflation.
In this way there is a truly flat direction 
along $\phi$ and the AD mechanism
can be implemented  as originally considered by Affleck and Dine. 
This argument is only exact at tree level. Strictly speaking, there
are small contributions to $m_\phi$  coming from higher loop corrections. 
These arise at two--loop or three--loop levels, depending on
 whether $\phi$ does or does not carry any  charge in common with the 
 U(1) charged fields \cite{Dv}. In addition, there are the 
expected $O($TeV) low-energy supersymmetry breaking contributions. 
In any case, $\phi$ will acquire a large VEV during inflation
due to quantum fluctuations, if the correlation length for de Sitter 
fluctuations $l_{coh} \simeq H^{-1}$exp(3$H^2/ 2 m_\phi^2)$
\cite{Bu}, is large compared to the horizon size.  In fact,
the present length corresponding to $l_{coh}$, namely
$l_{coh} (a_0/a_{infl})$ (where $(a_0/a_{infl})$ is the ratio of 
scale factors), should be larger than the horizon
size at present, $ct_0$. Using $H = 10^{13}$GeV, $t_0= 18$ sec
and assuming a radiation dominated universe ($a\sim t^{1/2}$), one
obtains the condition $H^2/m_\phi^2 \simgt 40$ \cite{DRT}
($\simgt 30$, if matter domination is assumed, since $a\sim t^{2/3}$),
which is easily fulfilled in this context.

\vspace{0.3cm}
\noindent
In conclusion, the AD mechanism can be implemented in a
 D--inflationary model and the preferred case is that  of $q_\phi=0$,
i.e. when the AD field(s) is (are) not charged under the relevant 
D--symmetry. Then
$m_\phi^2 \ll H^2$ and a large $\langle\phi\rangle_{in}$ is produced
by quantum fluctuations. The AD mechanism can thus  be implemented in
the original, ``old-fashioned", way. 
This scenario illustrates the fact that scalar fields do
not necessarily  get effective masses squared
(of either sign)
of $O(H)$ during inflation. Essentially vanishing masses are also
perfectly possible. Let us also emphasize that this scenario is quite
insensitive to the SUGRA details, in particular to the form
of the K{\"a}hler potential.

\section{F--Inflation}

In this section we will consider the case when inflation (and the
corresponding SUSY breaking) is driven by a non-vanishing F--term
of the appropriate size\footnote{F--inflation has been disputed 
(see e.g. Ref. \cite{BD}) because one naturally expects $O(H)$ effective
masses during inflation for all the scalars, which when applied to the
inflaton itself spoils the necessary slow rollover.  This is
not necessarily true.  One of the most interesting conclusions of this 
section and the following one is that very tiny masses are also possible,
 which makes
F--inflation as attractive, at least, as D--inflation.}.
Concerning the implementation of the AD mechanism in F--inflationary 
scenarios, the main question (see Sect. 1) is whether it is possible 
to get an effective mass squared $m_\phi^2<0$ 
or $m_\phi^2\simeq 0$ for the
AD field, $\phi$, during inflation. This will
automatically yield $\langle\phi \rangle_{in}\neq 0$, setting the onset
for subsequent baryogenesis, as explained in the Sect. 1. 
To answer this question, already addressed in Ref.\cite{DRT}, we need to 
 examine the effective potential,
$V$, in a SUGRA theory. Neglecting the contribution of the D--terms
(analyzed in the previous section) this is given by
\be
V= e^G\left( G_{\bar j} K^{\bar j l}G_{l}-3\right)
= F^{\bar l} K_{j \bar l}F^j-3e^G~.
\label{VFD}
\ee
Here $G=K+\log|W|^2$ where $W$ is the superpotential, $K^{\bar l j}$
is the inverse of the K{\"a}hler metric $K_{j \bar l}\equiv \partial K
/\partial z_j \partial \bar z_l$, $z_j$ 
are the (scalar components)
of the chiral superfields and $F^j=e^{G/2}K^{j\bar k}G_{\bar k}$ are the
corresponding auxiliary fields.
During inflation
\be
\langle V \rangle_{in} = V_0 \simeq H^2 M_P^2,
\label{Vi}
\ee
which implies that some $F$ fields are different form zero, thus
breaking SUSY. The effective gravitino mass squared 
during the inflationary 
epoch is given by  $m_{3/2}^2=e^{G}= e^{K}|W|^2$ in $M_P$
units. Notice that $V_0=F^{\bar l} K_{j \bar l}F^j-3m_{3/2}^2$, so,
unless there is some fine--tuning, $m_{3/2}^2$ is at most of $O(V_0)$,   
$m_{3/2}^2 \leq O(V_0)$. The SUSY breakdown
induces soft terms for all the scalars, in particular for $\phi$.
More precisely, the value of the effective mass squared, $m_\phi^2$, is
intimately related to the form of $K$. (As we will see shortly,
the form of $W$ is relevant for higher order terms, but not for
$m_\phi^2$.) It is convenient to parametrize $K$ as 
\be
K = K_0(I) + K_{\phi\bar\phi} |\phi|^2 + \cdots\;,
\label{K}
\ee
where $I$ represents generically the inflaton or inflatons\footnote{
If there are several AD fields which are mixed in the kinetic
term, $K_{\phi\bar\phi} |\phi|^2$, we are free to choose a basis of 
$\phi$--fields where it becomes diagonal.}.

\noindent Let us first show that it is {\em impossible} to get
the desired result, $m_\phi^2<0$ or $m_\phi^2\simeq 0$,  if
there is no mixing between $\phi$ and $I$ in the quadratic term of $K$,
 i.e. if
 $K_{\phi\bar\phi}\neq K_{\phi\bar\phi}(I)$.
To simplify the argument let us first assume that 
$W$ does not contain either
effective mass terms for $\phi$ or couplings between $\phi$ and $I$.
Then, it is clear from (\ref{VFD}) that a  mass term $m_\phi^2 |\phi|^2$
has two effective contributions 
\be
e^{K_0(I)}|W|^2 \left\{K_{\phi\bar\phi} |\phi|^2\right\} + 
e^{K_0(I)}|W|^2 K_{\phi\bar\phi} |\phi|^2 
\left\{G_{\bar j} K^{\bar j l}G_{l}-3\right\} \;,
\label{Vm2}
\ee
which lead to an effective mass squared
\be
m_\phi^2=m_{3/2}^2 + V_0/M_P^2\ ,
\label{mphi}
\ee
for the canonically normalized field $(K_{\phi\bar\phi})^{1/2}\phi$.
Hence $m_\phi^2$ is of $O(H^2)$ and positive and, therefore,  
$\langle\phi \rangle_{in}=0$ and the AD mechanism cannot be implemented.

Things may get more complicated if $W$ contains terms leading to
effective masses. Such  terms would generically read
\be
\lambda \eta_1\cdots \eta_n  \phi^m~~,
\label{Wphim}
\ee
where $\eta_j$ are fields that develop non-vanishing VEV's, at least
during inflation (the inflaton $I$ may be one of them), and $m=1,2$.
Let us continue to assume that the quadratic term of $K$ does not depend
on $\eta$, i.e.
$K_{\phi\bar\phi}\neq K_{\phi\bar\phi}(\eta)$. Then,
calling $\langle \lambda \eta_1\cdots \eta_n\rangle_{in}=M^{3-m}$,
besides the previously obtained positive mass terms ($\simeq H^2 
|\phi|^2$), we get positive  F--terms of order $M^{6-2m}|\phi|^{2m-2}$,
$M^{6-2m}|\langle \eta\rangle_{in}|^{-2}|\phi|^{2m}$, which for $m=1,2$ 
are large positive mass terms. Furthermore, there appear soft terms
of order $H M^{3-m} \phi^m$ which are effectively negative
 for an appropriate choice of the phases of the fields.
For $m=2$ these terms provide an effective negative mass squared
contribution. However, it
is unlikely that this would dominate over the previous positive 
contributions  of order $ H^2 |\phi|^2$ and $M^2|\phi|^2$, since
we expect 
either $MH \simlt M^2$ or $MH \simlt H^2$. Finally, for $m=1$ the effective 
linear term $\sim HM^2\phi$ guarantees $\langle\phi \rangle_{in}\neq 0$
(more precisely, $\langle\phi \rangle_{in}\simlt H$). We notice, 
nevertheless, that in this case again the flat direction of the
potential is lifted by inappropriate 
(quadratic in this case) terms $\sim |\phi|^2$.
As mentioned in the previous sections, according to Ref.\cite{DRT}
 this  implies that
the $\phi$--field does not follow the instantaneous minimum of $V$
after inflation, which may spoil the 
AD mechanism\footnote{This scenario, although rather artificial,
would deserve further analysis  anyway since, contrary to the scenarios
considered in Ref.\cite{DRT}, $\langle\phi\rangle$ is not
due to a negative quadratic  term in the potential, but to a linear one.}.
 
\vspace{0.3cm}
It is clear, therefore, that  a successful 
implementation of the AD mechanism in SUGRA theories and F--inflation
scenarios requires a mixing in the quadratic term of $K$
 of the inflaton $I$
 and AD fields $\phi$, i.e.
$K_{\phi\bar\phi} = K_{\phi\bar\phi}(I)$ in Eq.(\ref{K}).

\noindent Our next point is to show that this mixing should be
 remarkably strong.
Let us consider the following simple scenario
\be
K = K_0(I) + |\phi|^2 + a|I|^2 |\phi|^2 ,
\label{K2}
\ee
where $a$ is some unspecified coupling. This is the simplest
possible modification of minimal SUGRA to include the required mixing
between $I$ and $\phi$. Precisely a term  
$|I|^2 |\phi|^2$ in $K$ was invoked in Ref.\cite{DRT} to get negative
mass terms for $\phi$. Assuming for simplicity that there is a single 
inflaton $I$ with $|F_I|\neq 0$, we get from (\ref{VFD},\ref{Vi})
\bea
\langle V \rangle_{in} &=& \left(F^I\right) K_{I \bar I} 
\left(F^I\right)^*
-3e^{K_0}|W|^2
\nonumber \\
&=&
e^{K_0}
\left\{(W_I + K_I W) K^{\bar I I} (W_I + K_I W)^* -3|W|^2\right\}~.
\label{Vi2}
\eea
where $W_I =\partial W/\partial I$, $K_I= \partial K /\partial I$
and $W$ simply denotes $\langle W\rangle$.
Then, from the general expression for $V$, Eq.(\ref{VFD}), we
extract the various relevant contributions to the $\phi$ soft terms
\bea
&&e^{K_0 + K_{\phi \bar \phi}|\phi|^2}
\left\{ (W_I + K_I W) K^{\bar I I} (W_I + K_I W)^* -3|W|^2\right\},
\nonumber \\
&&e^{K_0}
\left\{ \left|W_\phi + K_\phi W\right|^2 
K^{\bar \phi \phi} -3 \left(W^* W(\phi)\;+\;{\rm h.c.}\right)\right\},
\nonumber \\ 
&&e^{K_0}
\left\{ (W_\phi + K_\phi W) K^{\bar I \phi} (W_I + K_I W)^* 
\;+\;{\rm h.c.}\right\}~~.
\label{mphi2}
\eea
The terms proportional to $W(\phi)$ give A--terms  of  $O(H)$ in the 
$\phi$ potential, while the terms proportional to $|\phi|^2$ give the
effective mass squared $m_\phi^2$.
 For instance, if $I$ is a canonically normalized field up to 
$O(|\phi|^2)$ contributions, i.e. if $(K_0)_{I\bar I}=1$, then
\be
m_\phi^2 = (1+a|I|^2)\left[
|F^I|^2\left(1-\frac{a}{(1+a|I|^2)^2}\right)-2m_{3/2}^2
\right]~~,
\label{mphi3}
\ee
where the requirements of a 
  positive kinetic energy for $\phi$ and  of a positive
cosmological constant $V_0$ yield, respectively, 
$1+a|I|^2>0$ and $|F^I|^2>3 m_{3/2}^2$. 
From (\ref{mphi3}) it is easy to determine the 
required value of $a$ in order to get $m_\phi^2\le 0$. 
In particular, for $|I|^2 > 3/4$ (in Planck units) 
there is no value
of $a$ for which $m_\phi^2\le 0$.
 This can actually be frequently the case,
 since one expects the inflaton to get VEV's of order $M_P$
during the inflationary process. For
 $|I|^2 \ll 1$, instead,  we can obtain negative 
  masses squared $m_\phi^2\le 0$ 
 if $a> 1/3$. This  quite strong mixing between the inflaton and the 
AD field is not at all trivial to obtain in models.
Still (as explained in Sect. 1), one has to assume the existence 
of F-contributions (preferably from non-renormalizable operators)
that eventually lift the flat direction of $V$ for large values
of $\phi$, which is not a strong requirement on models. 

To summarize the results of this section, in order to obtain  
$m_\phi^2\le 0$  during inflation, a strong mixing between the inflaton
and $\phi$ in the quadratic term ($\propto |\phi|^2$)
of $K$ is required. This is quite a strong constraint for SUGRA
scenarios, which excludes, in particular, minimal SUGRA.  

On the other hand,
the other desired possibility, that of a very small mass $m_\phi^2\sim 0$ 
(also welcome
for a successful inflation, as mentioned in footnote 4) does not seem 
natural at first sight. However, the study of the SUGRA
scenarios coming from strings provides beautiful surprises in this
sense, as we are about to see.

\section{String Scenarios}

Undoubtedly, the best motivated SUGRA scenarios are those coming
from string theories, which represent our best candidate for a
fundamental theory. Here we examine the capability of string scenarios
to implement AD baryogenesis.

As mentioned in Sect. 2 the D--inflation case is quite insensitive to 
the details of the particular SUGRA at hand. Correspondingly, all 
the results obtained there can be translated integrally to
string scenarios. Our only additional comment is that the existence
of an anomalous $U(1)$, which seems to facilitate the possibility
of D--inflation, is very common in string models, especially
in the most realistic ones \cite{CKM,CM2}.

Concerning F--inflation scenarios, string models do present special
characteristics since the K{\"a}hler potential, $K$, which plays
a crucial role, is greatly constrained.
We have seen in the previous section that the implementation of
the AD mechanism in this instance poses important restrictions on 
the form of $K$. So, we should first wonder whether the string 
K{\"a}hler potentials are able at all to accommodate AD baryogenesis.
If the answer is positive, we should then ask what particular string
models are favoured (or excluded) by this requirement.

In order to be concrete we will consider orbifold constructions,
which are very well known string models and, besides, are
extremely interesting for phenomenology. However, as it will become
clear, most of the conclusions are completely general.
The corresponding (tree--level\footnote{Perturbative corrections to
Eq.(\ref{Kor}) are known at one-loop level \cite{DKL} and are
small, so they do not affect any of the results presented here.
On the other hand, non-perturbative corrections are very poorly
known (see e.g. Ref. \cite{Ca} for an analysis of their possible
phenomenological significance).}) 
K{\"a}hler potential is given by \cite{DKL}
\be
K =  -\log(S+\bar S)-3\log(T+\bar T)+\sum_j(T+\bar T)^{n_j}|z_j|^2~.
\label{Kor}
\ee
Here $S$ is the dilaton and $T$ denotes generically the moduli 
fields\footnote{Eq.(\ref{Kor}) is written with the usual simplification
of considering a single ``overall modulus'' $T$.}, $z_j$ are the chiral
fields and $n_j$ the corresponding modular weights. The latter depend
on the type of orbifold considered and the twisted sector to which
the field belongs. The possible values of $n_j$  are 
$n_j= -1,-2,-3,-4,-5$. The discrete character of $n_j$ will play
a relevant role later on.

Since a strong mixing between the inflaton
and the AD field $\phi$ in the quadratic term 
 ($\propto |\phi|^2$) of $K$ is required (see Sect. 3), our first 
conclusion is that $T$ is the natural inflaton candidate
 in string theories. 
 
 This is a strong conclusion,
and  a very satisfactory one, since $T$ and also $S$ are, in fact,
 very suitable
candidates for inflatons for other reasons. Namely, $S$ and $T$
are present in all string constructions (at least in the 
perturbative approach) and their interactions with observable 
sector fields are gravitationally suppressed, as is expected for the 
inflaton. In addition, $S$ and $T$ have perturbatively flat 
potentials, which is appropriate for slow 
rollover\footnote{This can also be the origin of a cosmological
moduli (Polonyi) problem \cite{BKN}, but this is out of the
scope of the present work.}.
 Finally, we expect $\langle S\rangle, \langle T\rangle\simeq O(M_P)$
at low energy\footnote{ This comes from the fact that 
$\langle S\rangle$ and $\langle T\rangle$ have precise physical
meanings. Namely, $\langle S\rangle$ is the value of the unified
gauge coupling constant and $\langle T\rangle$ is the squared
radius of the compactified space, both in Planck units.}; so
some phase--transition-like process is expected for the $S$
and $T$ fields, which could well trigger inflation.
It should be noted that all the comments in this paragraph
are general for all string constructions.

We should recall, however, that a strong mixing in $K$ is a necessary
but not sufficient condition for $m_\phi^2\le 0$. We must then examine
the precise value of $m_\phi^2$ in the presence of a non-vanishing
cosmological constant $\langle V \rangle_{in}=V_0>0$. If inflation is
driven by the   $T$ (and/or $S$) fields, 
we expect non-vanishing $T$ (and/or $S$)
F--terms during inflation. Then, the corresponding soft terms, in
particular mass terms, for $\phi$ are straightforwardly extracted
from eqs.(\ref{VFD}, \ref{Kor}). This was precisely the sort of
scenario considered in Refs.\cite{KL2,BIM}.
Although the motivation of these works was different,
namely to study the form of the soft breaking terms at low--energy 
with $m_{3/2}=O(1$ TeV),  their results are applicable
here. The only difference here is that the scale of the breaking
is much higher and the non-vanishing cosmological constant,
$V_0>0$, plays a major role. In particular, the effective $\phi$ 
mass squared, $m_\phi^2$, which is especially relevant in our context,
is given by \cite{BIM}
\be
m_\phi^2 = m_{3/2}^2\left\{(3+n_\phi\cos^2\theta)C^2-2\right\}~,
\label{mphi4}
\ee
where $m_{3/2}^2=e^{K}|W|^2$, 
$\tan^2\theta= (K_{S \bar S}/
K_{T \bar T})\left|F^S / F^T\right|^2$, 
$C^2= 1+ [V_0 /(3M_P^2m_{3/2}^2)]$ and $n_\phi$
is the modular weight of the AD field $\phi$. Notice that $C^2>1$
and the condition for $m_\phi^2\le 0$ reads
\be
n_\phi\le
\frac{1}{\cos^2\theta}
\left( \frac{2}{C^2}-3 \right)~.
\label{cond1}
\ee
{}From (\ref{cond1}) we conclude the following.

\begin{description}

\item[{\em i)}]
If $\cos^2\theta=0$ ($S$--driven inflation), then $m_\phi^2
=m_{3/2}^2+V_0>0$, as expected.
Thus, as we  already concluded, 
``$S$--inflation'' is excluded by AD baryogenesis.

\item[{\em ii)}] 
If $\cos^2\theta=1$ ($T$--driven inflation), then $m_\phi^2\le 0$
is perfectly possible. In particular, it certainly occurs if 
$n_\phi\le -3$. In fact, states with $n_\phi\le -3$ occur in all
the orbifold  constructions, so $m_\phi^2\le 0$ is
perfectly natural in this context.

\end{description}

\noindent
The previous  conclusions {\em i)} and  {\em ii)} have been obtained
in the context of orbifolds, but are, in fact, much more general.
In particular, conclusion $i)$ is completely general since the
tree--level dependence of $K$ on $S$ is universal
in all string constructions. Conclusion $ii)$ is also very general
since the coupling of the chiral fields to the moduli in $K$ is
basically determined by modular invariance. We should mention
here that generic Calabi--Yau compactifications with large radius
($T$) have K{\"a}hler potentials as in eq.(\ref{Kor}), but with
all the fields in the untwisted sector ($n=-1$), so $m_\phi^2>0$
from (\ref{mphi4}). 
Therefore, those constructions are not favorable for AD baryogenesis. 
If the Calabi--Yau compactification is not in the large radius
limit, the corresponding expression
for $K$ is more involved and depends on the type of Calabi--Yau
compactification, so it would deserve a separate analysis.

\vspace{0.3cm}
\noindent
From $i)$ and $ii)$ we conclude that $T$--inflation is required 
for AD baryogenesis (a mix of $S$ and $T$--inflation could also work).
Hence, we will take in what follows
\be
\cos^2 \theta =1\ .
\label{costheta}
\ee
In this context we examine next, two particularly interesting limits 
that could well be realized in practice. Namely, since 
$V_0= K_{T \bar T}|F^T|^2-3m_{3/2}^2=3H^2 M_P^2$, it may
perfectly happen that $K_{T \bar T}|F^T|^2\gg m_{3/2}^2$, and 
thus $C^2\gg 1$ (see definition of $C^2$ after eq.(\ref{mphi4})). 
Another possibility is 
$K_{T \bar T}|F^T|^2 = O(m_{3/2}^2)$, and thus $C^2=O(1)$. Let us
analyze the two limits  separately. 

\begin{itemize}

\item
If $C^2\gg 1$, then, from eq.(\ref{mphi4})
\be
m_\phi^2\simeq H^2(3+n_\phi). 
\label{msmall}
\ee
Hence for $n_\phi=-3$ we get $m_\phi^2\simeq 0$.
So, we see that the possibility of a very small mass
$m_\phi^2\simeq 0$  can occur in 
F--inflation, as it was the case in D--inflation. Notice that
there is no fine-tuning here, since $n_\phi$ is a discrete number
which can only take the values $n_\phi=-1,-2,-3,-4,-5$.

This is also good news for F--inflation itself. As has been
pointed out in the literature (see footnote 4), F--inflation
has the problem that if the inflaton mass is $O(H)$, as expected
at first sight during inflation, then the necessary slow rollover
is disabled. We see here, however, that a hybrid--inflation
scenario \cite{Linde} in which $T$ is the field responsible for 
the large $V_0$ and a second field (any one with $n=-3$) is
responsible for the slow rollover is perfectly viable.

\item
If $C^2=O(1)$, the probability of obtaining $m_\phi^2<0$ is larger.
In particular, it is clear from (\ref{mphi4}) that
 if $C^2\le 2$, then $m_\phi^2\le 0$ whenever
$n_\phi\le -2$, which is a very common case.

\end{itemize}

\section{Conclusions}

Affleck--Dine baryogenesis requires the production of large 
expectation values for certain scalar fields (the AD fields)
at early (inflationary) times. This is only possible if the
effective soft masses squared coming from the inflationary breaking
of SUSY are negative or very small. This poses strong constraints
on the type of SUGRA theory. If inflation is driven by a D--term 
(D--inflation), the most favoured (and almost unique) scenario
occurs if the AD fields, $\phi$, are not charged under the 
gauge group of the relevant D--term. Then, $m_\phi^2\simeq 0$
and the AD mechanism occurs as originally considered by Affleck 
and Dine. This possibility
had been considered implausible in the recent literature. 
If inflation is driven by an F--term (F--inflation), then 
$m_\phi^2\le 0$ requires  a strong mixing of $\phi$ and the 
inflaton field
in the quadratic term ($\sim |\phi|^2$) of the K{\"a}hler potential. 
This (necessary but not sufficient) condition is not trivial and 
excludes in particular
minimal SUGRA. Amazingly, string-based SUGRA theories satisfy
the conditions for AD baryogenesis in a beautiful way, provided
that inflation is driven by a modulus ($T$) field. Then 
$m_\phi^2\le 0$ for certain values of the $\phi$ modular weight
(in particular $n_\phi\le -3$ always works). In addition,
the  possibility of $m_\phi^2\simeq 0$ can also appear in a natural
way, with no fine-tuning at all, thanks to the discrete character
of $n_\phi$. This is a nice result which also shows,
as explained after Eq.(\ref{msmall}), that 
$T-$ hybrid inflation is viable in string-based SUGRA scenarios.

\section*{Acknowledgements}
We thank the Aspen Center for Physics, where this paper was initiated,
for its hospitality. We also thank M. Quir{\'o}s for a careful
reading of the paper.



\def\MPL #1 #2 #3 {{\em Mod.~Phys.~Lett.}~{\bf#1}\ (#2) #3 }
\def\NPB #1 #2 #3 {{\em Nucl.~Phys.}~{\bf B#1}\ (#2) #3 }
\def\PLB #1 #2 #3 {{\em Phys.~Lett.}~{\bf B#1}\ (#2) #3 }
\def\PR  #1 #2 #3 {{\em Phys.~Rep.}~{\bf#1}\ (#2) #3 }
\def\PRD #1 #2 #3 {{\em Phys.~Rev.}~{\bf D#1}\ (#2) #3 }
\def\PRL #1 #2 #3 {{\em Phys.~Rev.~Lett.}~{\bf#1}\ (#2) #3 }
\def\PTP #1 #2 #3 {{\em Prog.~Theor.~Phys.}~{\bf#1}\ (#2) #3 }
\def\RMP #1 #2 #3 {{\em Rev.~Mod.~Phys.}~{\bf#1}\ (#2) #3 }
\def\ZPC #1 #2 #3 {{\em Z.~Phys.}~{\bf C#1}\ (#2) #3 }

\end{document}